# DNA energy constraints shape biological evolutionary trajectories.

Piero Fariselli[1*], Cristian Taccioli[2*], Luca Pagani[3,4], and Amos Maritan[5]

*[1]Dipartimento di Scienze Biomediche, Università di Torino, Via Santena 19, Torino, 10126, IT. [2]Dipartimento di Medicina Animale, Produzione e Salute, Università di Padova, Viale dell'Università 16, Legnaro (Padova), 35020, IT. [3]Dipartimento di Biologia, Università di Padova, Via Ugo Bassi 58/B, Padova, 35131, IT. [4]Estonian Biocentre, Institute of Genomics, University of Tartu, Riia 23b, Tartu, 51010, EE. [5]Dipartimento di Fisica e Astronomia, G. Galilei, INFN, Università di Padova, Via Marzolo 8, Padova, 35131, IT.*

**These authors contributed equally to this work.*

**Most living systems rely on double-stranded DNA (dsDNA) to store their genetic information and perpetuate themselves. This biological information has been considered the main target of evolution. However, here we show that symmetries and patterns in the dsDNA sequence can emerge from the physical peculiarities of the dsDNA molecule itself and the maximum entropy principle alone, rather than from biological or environmental evolutionary pressure. The randomness justifies the human codon biases and context-dependent mutation patterns in human populations. Thus, the DNA "exceptional symmetries", emerged from the randomness, have to be taken into account when looking for the DNA encoded information. Our results suggest that the double helix energy constraints and, more generally, the physical properties of the dsDNA are the hard drivers of the overall DNA sequence architecture, whereas the biological selective processes act as soft drivers, which only under extraordinary circumstances overtake the overall entropy content of the genome.**

## INTRODUCTION

The biological information contained within a dsDNA molecule, in terms of a linear sequence of nucleotides, has been traditionally considered the main target of selective pressures and neutral drift(1-3). However, in this information-centred perspective, certain emerging traits of the genetic code, such as symmetries between nucleotides abundance(4-7), codon preferences(8,9), and context-dependent mutation pattern(10), are difficult to explain. In 1950, Erwin Chargaff made the important observation that the four nucleotides composing a double helix of DNA (Adenine, A; Cytosine, C; Guanine, G and Thymine, T) are symmetrically abundant(11) (number of A = number of T and number of C = number of G). This symmetry, named Chargaff's first parity rule, played a crucial role in the discovery, in 1953, of the double helix structure of DNA(12-14). In 1968 Chargaff extended his original observation into the Chargaff's second parity rule(15-17), which states that the same sets of identities found for a double helix DNA also hold on every single strand of the same molecule. In other words, in every single strand of a dsDNA molecule, the number of adenines is almost equal to the number of thymines and the number of guanines is almost equal to the number of cytosines. This rule does not hold for single-stranded DNA (ssDNA), and it has been found to be globally valid for all the dsDNA genomes with the exception of mitochondria(18,19). An updated confirmation of these previous observations is reported in Figure 1 and Table 1S.

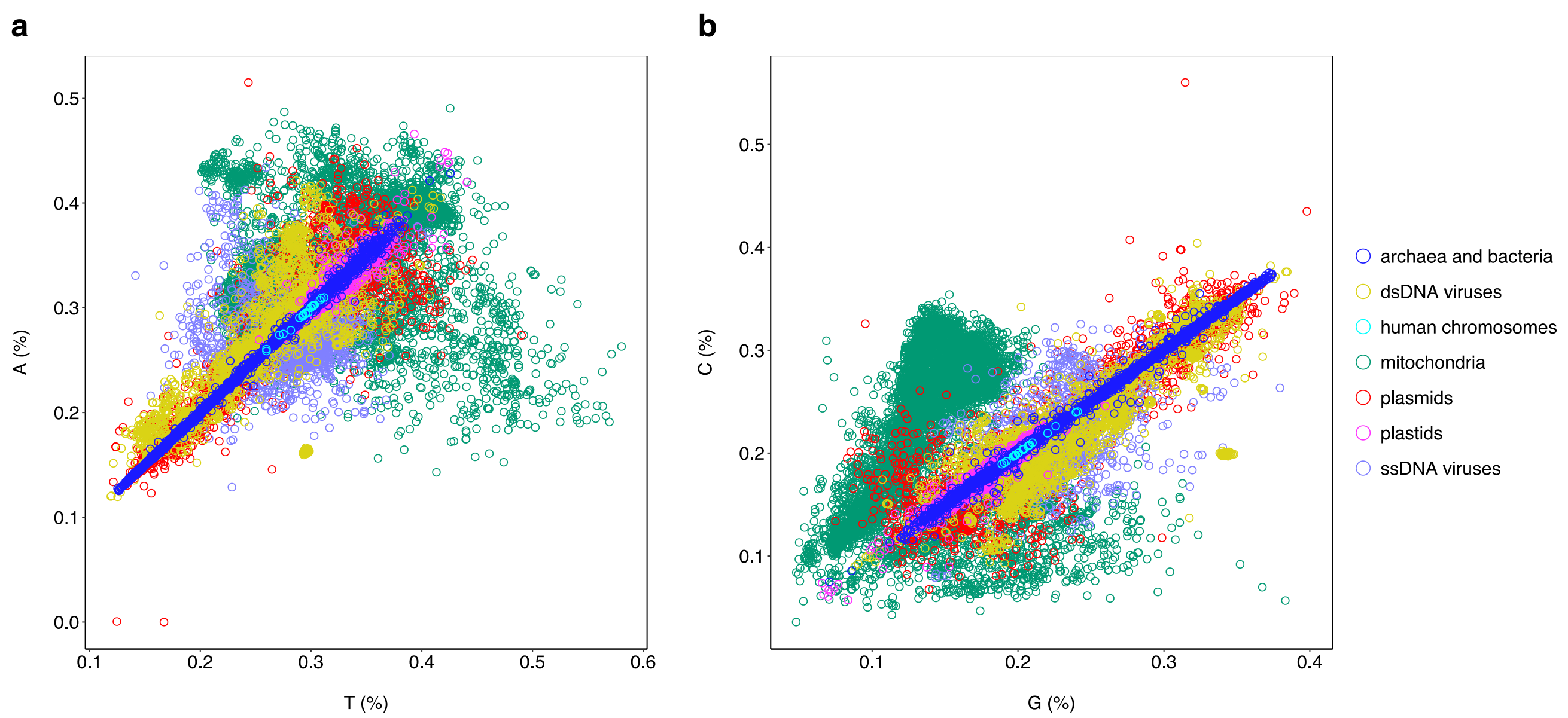

**Figure 1 | Percentage of Adenine (A) versus Thymine (T) and Cytosine (C) versus guanine (G) computed for different organism genomes. a**, Percentage scatterplot of Adenine versus Thymine. Archaea, bacteria and human chromosomes have the highest Pearson correlation values ($R^2$=0.99). Plastids, plasmids and double-stranded DNA (dsDNA) viruses have a Pearson correlation $R^2$ greater than 0.6. Mitochondria and single-stranded DNA (ssDNA) viruses do not show a significant correlation ($R^2$ < 0.04). **b,** A similar graph was obtained plotting the percentage of Cytosine (C) versus the percentage of Guanine (G) using the same set of organism genomes. Archaea, bacteria and human chromosomes have the highest Pearson correlation values ($R^2$=0.99). Plastids, plasmids and double-stranded DNA (dsDNA) viruses have a Pearson correlation $R^2$ greater than 0.7. Mitochondria and single-stranded DNA (ssDNA) viruses do not show a significant correlation ($R^2$ < 0.04).

Chargaff's second parity rule has previously been extended to all the possible k-mers up to 10 bases(4,5) within a dsDNA molecule and holds only for k-mers and their reverse complements (here named "RC-pairs"), but not for any alternative permutation of the reverse complement k-mers. This symmetry, which holds for all the dsDNA genomes, has been recently named "exceptional symmetry"(6). As an example of this, here we consider the complement pairs ("C-pairs"). In this case for example (see Figure 2), the occurrences of the nucleotide $^{5'}$TTACG$^{3'}$ and its reverse complement sequence $^{5'}$CGTAA$^{3'}$ (are RC-pairs) in a single strand of a dsDNA genome are almost the same. Conversely, the frequencies of the C-pairs $^{5'}$TTACG$^{3'}$ and $^{5'}$AATGC$^{3'}$ in the same strand may differ significantly. Notice that the direction 5' -> 3' is conserved between RC-pairs (see Figure 1a), whereas it is inverted in C-pairs (see Figure 1b).

After fifty years from the discovery of Chargaff's second parity rule, there is not a generally accepted justification for its emergence, although several explanations have been proposed based on different models and hypothesis, such as: statistical(7,20,21), stem-loops(22), tandem duplications(23), duplication followed by inversions(24), inverted transpositions(25,26), and non-uniform substitutions(27). Notably, these explanations, although very promising, do not have predictive power, i.e. from them it is difficult to deduce testable predictions.

Here we hypothesize that the leading force shaping the DNA sequence in the genomes is the entropy and that the major cause of all these symmetries is the randomness. However, randomness does not imply uniformity and equality. As an example, the "random" process of blindly throwing stones in a rugged landscape generates a higher probability of finding the stones in the valleys than on top of the hills.

## RESULTS

### DNA symmetries from randomness

To find the probability distribution of the dsDNA sequences, we applied the maximum entropy approach (28), taking into account the energy constraints dictated by the DNA double helix structure. This is equivalent to finding DNA sequence arrangements corresponding to the minimum free-energy(29). According to this principle, the distribution maximizing the entropy is the least biased, among the ones satisfying the energy constraints. By using the probability distribution with the highest entropy, we are choosing the model that needs the smallest amount of information to be explained.

The double-energy interaction and its intrinsic symmetry (due to the Newton's third law) shifts the probability away from the uniform distribution. Given a dsDNA with the two strands $X$ (the plus strand) and $\bar{X}$ (the minus strand) the interaction energy $H(X,\bar{X})$ is equal to $H(\bar{X},X)$. This implies that the energy does not change if $X$ and $\bar{X}$ are interchanged, that is $H(X,\bar{X}) = H(\bar{X},X)$ (Figure 2a). As a consequence the probabilities of a sequence and its reverse complement are equal, $P(X) = P(\bar{X})$ (eq.(2), see methods).The most relevant fact is that the energy symmetry does not hold when comparing the interaction energy $H(X,\bar{X})$ with that of the complement sequence $X^c$ (its C-pair) on plus strand and its base-pairing sequence on minus strand $H(X,\bar{X}) \neq H(X^c,\bar{X}^c)$ (Figure 2b). This implies that, in general, $P(X) \neq P(X^c)$.

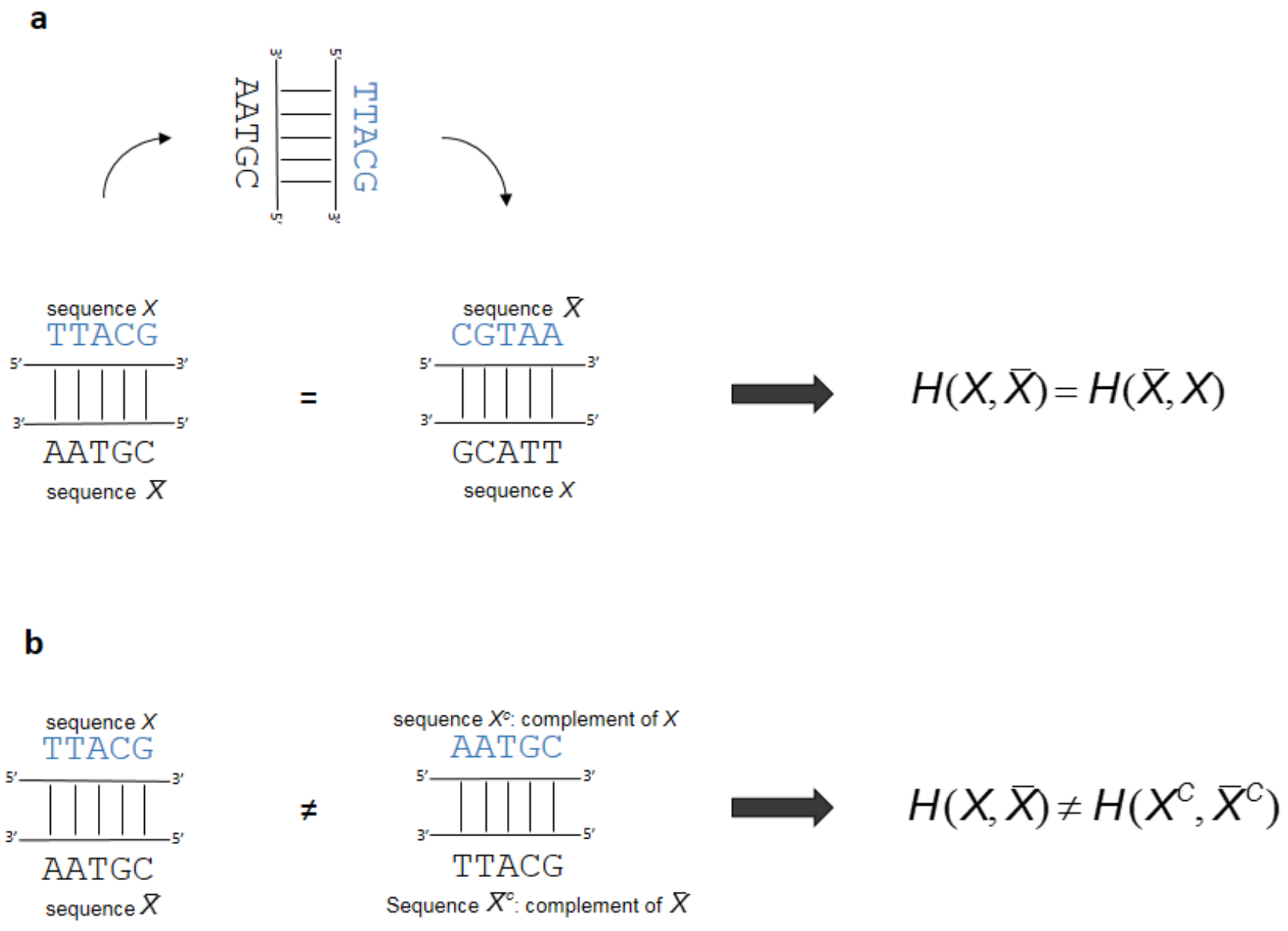

**Figure 2 | Energy symmetry of dsDNA**. **Scheme describing the interaction energy in RC- versus C-pairs. a**, The interaction energy between the sequence $X$ ($^{5'}$…TTACG…$^{3'}$) on plus strand and the sequence ($^{3'}$ …AATGC…$^{5'}$) $\bar{X}$ on the minus strand is the same of its reverse complement sequence $\bar{X}$ ($^{5'}$ …CGTAA…$^{3'}$) on the plus strand when interacting with the sequence $X$ ($^{3'}$...GCATT…$^{5'}$) on the minus strand. Thus, in RC-pairs this implies that $H(X,\bar{X}) = H(\bar{X},X)$. **b,** The energy equality does not hold when taking into account the interaction energy between the complement, $X^c$ ($^{5'}$…AATGC…$^{3'}$), of the original sequence $X$ on the plus strand, with its reverse complement $\bar{X}^c$ ($^{3'}$…TTACG…$^{5'}$) on the minus strand, that is $H(X,\bar{X}) \neq H(X^C,\bar{X}^C)$. Thus, this energy symmetry leads to the prediction that that the probability of finding a specific k-mer on a strand is the same of finding its reverse complement k-mer on the same strand, when we apply the energy constraint to a maximum entropy approach. In particular, this leads to Chargaff's second parity rule because of the sequences $X$ and $\bar{X}$ contain the same number of A-T and C-G. Note that in RC-pairs **(a)** that direction 5' -> 3' is conserved between the sequences TTACG and GCATT (the same is true for AATGC and CGTAA), whereas in C-pairs **(b)** the direction of the same sequences are inverted. In other words, the sequences in **(a)** are the same, while in **(b)** are specular.

The main prediction deducible from the equality $P(X) = P(\bar{X})$ (equation (2), see methods) is that the expected number of RC-pairs are equally balanced. This is only technically correct for infinitely long dsDNA sequences, and we can foresee deviations when the genome size decreases (as in the case of the DNA of viruses and organelles).

The maximum entropy solution predicts that the maximum length of a RC-pairs in a genome is not constrained to any specific length of a k-mer and depends only on the range of the energy interaction, which could in principle span the whole dsDNA genome. Therefore, our framework represents a generalization of the preliminary observations of an "exceptional symmetry"(6) to a more universal principle that here we call "Generalized Chargaff's Theory" (GCT) and which is built only upon a physical approach.

In this context, Chargaff's second parity rule is deduced from the maximum entropy and represents just a special case of GCT corresponding to *k*=1 (k-mer of length equal to 1). On the other hand, the inequality $P(X) \neq P(X^c)$ predicts that the frequencies of a k-mer and its complement are not correlated.

When the maximum entropy is applied to ssDNA (see supplementary materials), in the absence of any other known single-stranded energy interaction constraints, no correlation between the abundances of A and T (and between C and G, or of any other RC-pair) in ssDNA is expected. This second prediction of the model is consistent with the ssDNA virus data (Figure 1).

A third prediction of our physical formulation of GCT (equation (3), see methods) is the freedom for relative positions of RC-pairs in the dsDNA sequences, unlike alternative biological models that imply duplications with inversions and other local phenomena of DNA rearrangements, where relative positions of the two sequence of a RC-pair may be constrained by the duplication mechanism. According to our maximum entropy solution, the probability of finding a RC-pair in a double helix sequence is determined by the range of the interaction energy $H(X, \bar{X})$, and not necessary nearby within a dsDNA genome. This is confirmed by analysing a set of dsDNA genomes of several species, where the data show that the energy interactions span very distant sequence positions (see Supplementary Information and Figures 1S, 2S).

A fourth prediction of GCT concerns the empirical observation that RC-pairs are extremely frequent in dsDNA genomes (see Figure 3), while any other kind of pairs, such as C-pairs, are expected to exponentially decrease with their length. Indeed, a prediction of the expected number of RC-pairs and C-pairs, as a function of k (length of a k-mer), can be analytically derived (equations (4) and (5), see methods) and tested against experimental observations. Figure 3a shows an almost perfect, k-independent match between the predicted and observed DNA sequences in the Escherichia coli genome, taken as representative genome for prokaryotes.

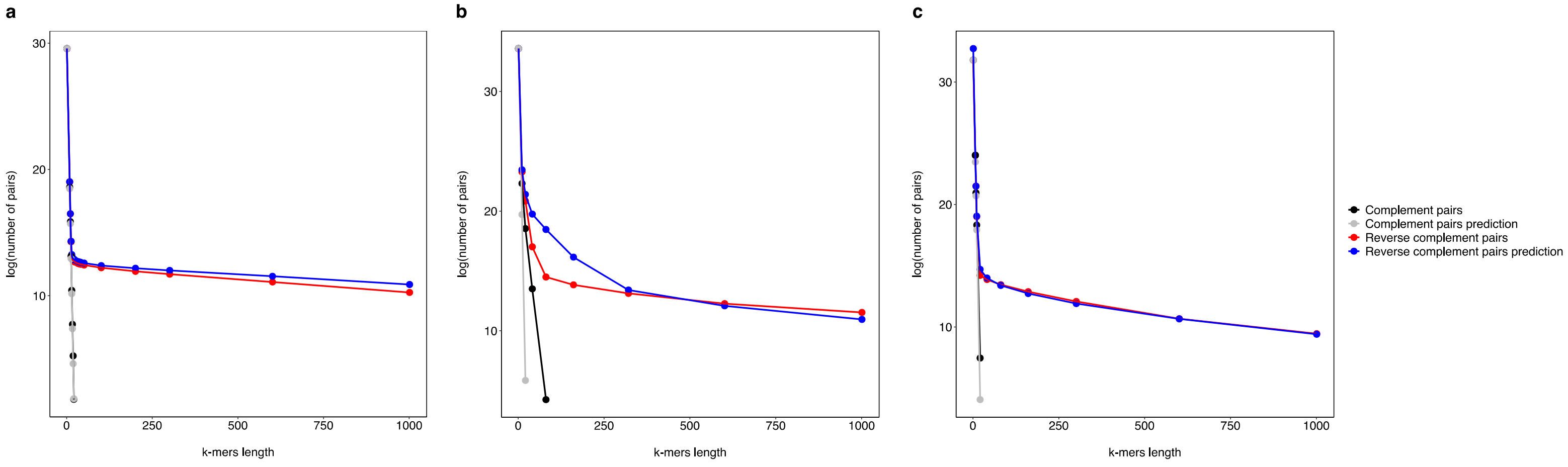

**Figure 3 | Predictions of the number of RC- and C-pairs in prokaryotic and eukaryotic genomes.** The logarithm of the number of RC- and C-pairs are plotted as a function of the k-mer length. The plot shows that RC-pairs are much more frequent compared to C-pairs. Moreover, the predicted and observed number of RC-pairs (blue and red dots respectively) decreases sub-linearly whereas the predicted and observed number of C-pairs (grey and black dots respectively) decreases exponentially (linearly in the graph). **a)** The predicted number of RC- and C-pairs perfectly match the observed data in Escherichia coli K12 genome**. b)** Human chromosome 22. The predicted number of RC and C-pairs are in agreement with the empirical data, but not at the same degree as for Escherichia coli **(a)**. **c)** Human

chromosome 22 with removed repeated regions. Now the match between our predictions and empirical data is of the same quality as for prokaryotes **(a)**.

When a eukaryotic genome is considered, here represented by the human chromosome 22 (the smallest in terms of nucleotides abundance), we find that the predictions of our equations (equations (4) and (5), see methods) are still in good agreement with the empirical data (see Figure 3b), but not at the same degree as for prokaryotes. However, when repeated regions (transposons, tandem repeats, low complexity regions, etc.) are removed, we found an optimal improvement between predictions and observations (Figure 3c).

This finding suggests the actual presence of two distinct portions within eukaryotic genomes: i) a stable core genome, similar to the prokaryotic DNA, which is at the GCT equilibrium, ii) regions originated from recent rearrangements events, that are either still evolving towards the equilibrium or are kept away from it by biological selective pressures. In evolutionary terms, we can imagine that, among all the chromosomal rearrangements affecting a genome, the ones that maintain or facilitate the emergence of RC-pairs are those that help the dsDNA genome to reach the equilibrium (maximum entropy or minimum free energy) and are hence positively selected in light of their energy balance rather than just for their biological information content.

Among the previously hypothesized mechanisms, the duplications followed by inversions (25-26), which create reverse complement sequences and led to GCT, are the most probable outcome in terms of genomic thermodynamic equilibrium. In this energetic view, the observed difference in abundance of various k-mers, can be interpreted as a result of the difference of free energy of the relative nucleotide sequences.

### Codon usage and evolutionary patterns in human population

In virtue of GCT we can also observe how biological events (such as duplications contained in the repeated tracts of the human genome) might generate detectable deviations from the maximum entropy solution. We focused on coding regions, which sequences are constrained by the fact that, under certain reading frames, tri-nucleotides are translated into amino acids following a species-specific code. Here we investigate whether, even in a given codon bias system, the GCT is still detectable as an entropic tendency. Through the human codon usage, we found a significantly positive correlation between the frequency of the codons and the frequency of the corresponding reverse-complement codons (Pearson correlation=0.5 with a p-value $< 10^{-4}$). On the contrary, there is no correlation between the codon frequency and those of their complement codons (Pearson correlation $R^2 = 0.0$ with a p-value $\approx 1$). This indicates that, despite biological selective pressures, there is a significant GCT trace in the human codon bias. We went further on this path, and we generated $10^7$ random permutations of the observed codon usage frequencies to evaluate the stability of the GCT signal in the human codon usage. The $10^7$ random codon frequencies preserve the amino acid abundance to maintain the same protein composition for each randomly generated codon usage (see methods). We then evaluate the Pearson correlation of the codon frequency as a function of the distance between the "true" human codon usage and the corresponding random permutations. The results, reported in Figure 4, show that the more the simulated codon usage is distant from the true one, the less (on average) the RC pairs correlate. Conversely, this does not happen for the C-pairs. The result confirms that the GCT signal in the human codon bias is robust. In broader terms, assuming "perfect GCT compliance" as the energy equilibrium, we could tentatively see the energy needed to deviate from the equilibrium while keeping a certain level of GCT "unbalance" (i.e. and $R^2$ of 0.5) within the adopted codon usage, as an upper limit for the evolutionary "energy gain" represented by the usage of alternative codon usage.

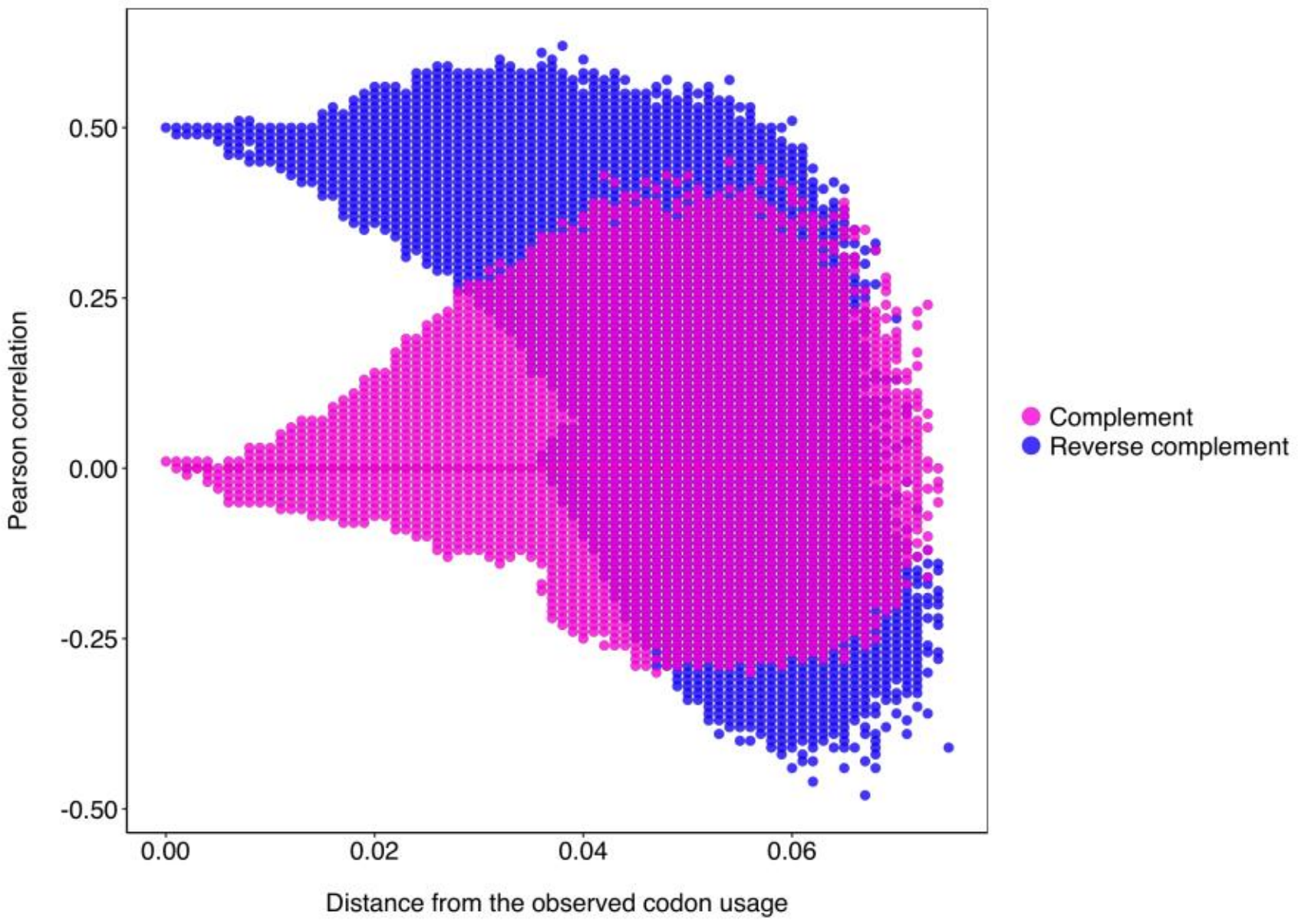

**Figure 4 | Pearson correlation of the codon frequency as a function of the distance between the observed human codon usage and random permutations.** The correlation is computed for the pairs of different codon frequency that are the reverse complement (RC-pairs) and simple complement (C-pairs). We generated $10^7$ random samples by shuffling the codon frequencies inside the group coding for the same amino acid, to keep constant the natural amino acid abundance. The $R^2$ between the computed Pearson (y-axis) and the distance (x-axis) for the RC-pairs is 0.65 ($R^2$ =0 for the C-pairs). This indicates that there is a tendency to lose the Chargaff's fingerprinting moving away from the true codon usage.

Remembering that in the human genome the abundance of trimers is not uniform and follows the GCT (see Supplementary Information Figure 3S), another validation of the role played by the dsDNA energy constraints is provided by the observation of context-dependent mutations within the human genomes of the human 1000 Genomes Project(30). As recently reported(10), when stratifying the occurrences of a given mutation A -> B within a human population by considering the 5' and 3' context nucleotides, the relative abundance of the various N**A**N -> N**B**N trimers (k=3) is not uniformly distributed (N can be any of the four DNA nucleotides for two fixed nucleotides A and B). We can assume that, among other causes, GCT, as a major driver toward the equilibrium of the genome, may be one of the reasons for the observed non-uniformity of the mutation N**A**N -> N**B**N. If this is the case, we expect that at population level, the evolutionary success of a given mutation type, approximated by its average population frequency should match the average population frequency of the mutation type substituting the RC of N**A**N into the RC of N**B**N, hence ensuring compliance with GCT (see Supplementary Information for further details). This is what we observe from empirical data (Figure 4Sb).

Overall, our results show that processes that increase the entropy of a double-stranded DNA molecule are favoured, and we speculate that exceptions to this trend may provide future opportunities to measure the energetic content of the biological information embedded in dsDNA sequences shaped by the natural selection.

## DISCUSSION

Here we showed that the intra-strand symmetries in the dsDNA emerge from the double-helix structure and the randomness. It appears that Nature does not fight against the entropy when not strictly necessary, even when information of a whole organism must be encoded in a DNA molecule. In this respect, we defined the principle of the Generalized Chargaff's Theory (GCT), which expands the unexpected "exceptional symmetry" found in double helix DNA as just the most likely and simplest probability distribution attainable for a duplex DNA. It is worth noticing that we obtained this result by solely imposing the base-pairing energy constraint, due by the symmetrical nature of the DNA double helix, and using the entropy maximization without specifying an energy form at quantitative level. These two simple physical ingredients, rather than biological events, seem to be at the basis of the observed genome-wide patterns that generalize the discovery made by Chargaff in 1968 (Chargaff's second parity rule).

The maximum entropy solution, furnishes support to the empirically proposed mechanism of the DNA inversions (25-26) as major driver toward the equilibrium and the creation of symmetries in the DNA. This is notably true also for the encoded genetic information of the species, since many local exceptions are not capable of altering the genome-wide picture as a whole. The maximum entropy fingerprint is also visible in the pattern of the human evolution, where the context-dependent mutations exhibit the same frequency for mutations in trimers and their reverse mutation in the RC-trimers.

## METHODS

**Maximum entropy and DNA energy constraint**. Assuming that the dsDNA is at the equilibrium, and that the DNA sequence is sufficiently long, we can estimate the probability of a sequence $P(X)$ using a maximum entropy approach considering that the average energy of the double helix interaction is fixed (see supplementary materials for more details). This means to maximize the following functional (with respect to $P(X)$):

$$F = -\sum_X P(X)\ln P(X) - \lambda \sum_X P(X) - \beta \sum_X \widehat{H}(X)P(X) \quad (1)$$

where for compactness we set $\widehat{H}(X) = H(X,\overline{X})$. The solution of this variational problem leads to the fact that the probability of a sequence $X$ is the same as its reverse complement sequence $\overline{X}$, i.e.

$$P(X) = \frac{1}{Z}e^{-\beta\widehat{H}(X)} = \frac{1}{Z}e^{-\beta\widehat{H}(\overline{X})} = P(\overline{X}) \quad (2)$$

because of the energy symmetry $\widehat{H}(X) = H\left(X,\overline{X}\right) = H\left(\overline{X},X\right) = \widehat{H}(\overline{X})$

From the result equation (2) it is possible to show (see supplementary materials) that the expected number of k-mer *w, n(w),* is the same of its reverse complement irrespectively of its length, or more formally:

$$\langle n(w)\rangle = \langle n(\overline{w})\rangle \quad (3)$$

The original Chargaff's second parity rule, is just the case of k-mer of length 1 ($w \epsilon \{A, C, G, T\}$). With this model, we can make some predictions about the expected number of RC- and C-pairs in a dsDNA long enough sequence. In particular, under some reasonable assumptions (see supplementary materials), for the expected number of C-pairs of length *k*, $n(w, w^c, k)$, we have:

$$\langle n(w, w^c, k)\rangle \simeq \frac{1}{2}n(k)^2 \cdot 4^{-k} \quad (4)$$

where $n(k) = N - k + 1$ is the number of possible k-mer in the dsDNA sequence of length *N*. Similarly, the expected number of RC-pairs is:

$$\langle n(w, \overline{w}, k)\rangle \simeq \frac{1}{2}\left(\sum_{i=1}^{n(k)}\langle n(w_i)\rangle - n(k)\right) \quad (5)$$

where $\langle n(w_i)\rangle$ is the expected number of k-mer equals to the one starting at the $i^{th}$ position of the DNA sequence and of length k. In this case we exploit equation (3) to evaluate the number of reverse complement k-mers (see supplementary materials).

**Codon usage analysis**. Each codon in the human genome has a reported frequency (codon usage frequency). We evaluated the Pearson correlation between the frequencies of each codon with respect to the frequencies of the corresponding Reverse-Complement codons (or Complement codons). This is done by creating a list V, whose elements are the frequency of the alphabetically ordered codons. Then we generated two other lists $V^{RC}$ and $V^{C}$, whose elements are the frequencies of the reverse complement codons and complement codons of V, respectively. Then, we computed the Pearson correlation between V and $V^{RC}$, and between V and $V^{C}$. Furthermore, we generated $10^7$ different random codon list R, by shuffling the original frequencies in V (by shuffling the elements $V_i$ of V). However, in order to keep the same amino acid frequency, and avoiding improbable amino acid frequencies in the genome, the shuffling is performed only inside the groups of the codons that code for the same residue (or stop codons). The number of possibility is still extremely high, due to the degeneracy of the code. For each new random codon R we computed as before, the Person correlation between R and the corresponding Reverse-Complement $R^{RC}$ and Complement $R^{C}$ lists. At the same time we measure the Euclidean distance $D(V,R)$ between the original V and the new random shuffled R as:

$$D(V,R) = \sqrt{\sum_i (V_i - R_i)^2} \qquad (6)$$

In Figure 4, we plot the Pearson correlation between R and $R^{RC}$ (or $R^{C}$) as a function of $D(V,R)$.

**Acknowledgements**

We thank T. Bobbo, A. Boattini and C. Vischioni for critical reading of our manuscript. P.F. was supported by the Italian Ministry for Education, University and Research under the programme "Dipartimenti di Eccellenza 2018-2022", Project code D15D18000410001. C.T. was supported by the University of Padova through M.A.P.S department under the programme "SID2017 Project code BIRD171214". L.P. was supported by the European Union through the "European Regional Development Fund", Project No. 2014-2020.4.01.16-0024, MOBTT53. A.M. was supported by "Excellence Project 2017" of the Cariparo Foundation.

**Author contributions**

P.F., C.T. and A.M. conceived the study. P.F. C.T and L.P run the computations. All the authors analyzed the data and wrote the manuscript.

**Competing interests**

The authors declare no conflict of interest.

**Supplementary Figures and Tables**

**Table S1**. Frequency of nucleotides in different groups of organism genomes.

**Figure 1S**. Scatterplot of bacteria k-mers vs their reverse complement sequences

**Figure 2S**. Scatterplot of k-mers against their reverse complement in six eukaryotic chromosomes with and without repeated regions.

**Figure 3S**. Trimer as a function of their frequency (y-axis) in the human genome

**Figure 4S**. Context dependent substitutions within the human genome (1000 Genomes Project European samples).

*Supplementary Information*

# DNA energy constraints shape biological evolutionary trajectories.

Piero Fariselli[1*], Cristian Taccioli[2*], Luca Pagani[3,4], and Amos Maritan[5]

[1]*Dipartimento di Scienze Biomediche, Università di Torino, Via Santena 19, Torino, 10126, IT.*
[2]*Dipartimento di Medicina Animale, Produzione e Salute, Università di Padova, Viale dell'Università 16, Legnaro (Padova), 35020, IT.*
[3]*Dipartimento di Biologia, Università di Padova, Via Ugo Bassi 58/B, Padova, 35131, IT.*
[4]*Estonian Biocentre, Institute of Genomics, University of Tartu, Riia 23b, Tartu, 51010, EE.*
[5]*Dipartimento di Fisica e Astronomia, G. Galilei, INFN, Università di Padova, Via Marzolo 8, Padova, 35131, IT.*

**These authors contributed equally to this work.*

# 1. MATEMATHICAL METHODS

## 1.1 Maximum Entropy Principle and the Constraint of the DNA Double Helix Interactions

For the sake of compactness, we introduce the following notation. We define a DNA sequence of length $N$ the string $X = a_1 \ldots a_N$, where $a_i$ is one of the possible four nucleotides {A,C,G,T}. We define two functions of a DNA sequence $X$: 1) the reverse sequence $X^r = a_N a_{N-1} \ldots a_1$, and 2) the complement $X^c = a_1^c a_2^c \ldots a_N^c$, where $a_i^c$ represent the complement bases, such as $A^c$ = T and $C^c$ = G. Finally, the reverse complement of a sequence $X$ is defined as $\overline{X} = (X^r)^c = (X^c)^r = a_N^c a_{N-1}^c \ldots a_1^c = \overline{a}_N \overline{a}_{N-1 \ldots} a_1$. The last equality comes from the fact that the reverse complement of a single nucleotide is just its complement. When we consider a dsDNA (double-stranded DNA), the two paired DNA $X$ and $\overline{X}$ interact with an energy $\widehat{H}(X) = H(X, \overline{X})$, which is symmetric under the exchange of $X$ with $\overline{X}$, i.e $\widehat{H}(X) = H(X, \overline{X}) = H(\overline{X}, X) = \widehat{H}(\overline{X})$, since the force exerted by $X$ on $\overline{X}$ is the same as the force that $\overline{X}$ exerts on $X$, and its strength depends on the specific DNA sequence $X$. On the other hand $\widehat{H}(X) = H\left(X, \overline{X}\right) \neq H\left(X^c, \overline{X^c}\right) = \widehat{H}(X^c)$. We then can define an average energy as:

$$\langle E \rangle = \sum_X \widehat{H}(X) P(X) = \sum_X H(X, \overline{X}) P(X) \tag{1}$$

where $P(X)$ is the probability of occurrence of the DNA strand $X$ and the sum is over all possible DNA sequences. Our purpose is to find the most probable and less informative distribution among the ones satisfying the constraint in eq. (1). This is achieved by maximizing the information entropy[1] $S$:

$$S = -\sum_X P(X) \ln(P(X)) \tag{2}$$

with respect to $P(X)$ with the constraints given by eq. (1) and by the normalization condition $\sum_X P(X) = 1$. By introducing the appropriate Lagrange multipliers, $\lambda$ and $\beta$, the function to maximize is:

$$F = -\sum_X P(X) \ln\big(P(X)\big) - \lambda \sum_X P(X) - \beta \sum_X \widehat{H}(X) P(X) \tag{3}$$

After the maximization of $F$, the probability can be written as:

$$P(X) = \frac{e^{-\beta \widehat{H}(X)}}{Z} = \frac{e^{-\beta H(X, \overline{X})}}{Z} \tag{4}$$

Where the constant Z is the partition function:

$$Z = \sum_X \exp\{-\beta \widehat{H}(X)\} \tag{5}$$

The Lagrange multiplier $\beta$ is related to $E$ in eq. (1) by the relation:

$$E = -\frac{\partial lnZ}{\partial \beta} \tag{6}$$

From the symmetry of the interaction energy ($\widehat{H}(X) = H(X, \overline{X}) = H(\overline{X}, X) = \widehat{H}(\overline{X})$), it follows that:

$$P(X) = \frac{1}{Z} e^{-\beta \widehat{H}(X)} = \frac{1}{Z} e^{-\beta \widehat{H}(\overline{X})} = P(\overline{X}) \tag{7}$$

This indicates that the probability of the existence of a DNA sequence in a genome, for example, is equal to the probability of its reverse complement.

## 1.2 Chargaff's Second Parity Rule from Maximum Entropy Principle

We will now show that the symmetry of the energy and the double helix interaction jointly with the maximum entropy principle is the origin of GCT (Generalized Chargaff's Theory): in a long-enough duplex DNA, the occurrence of a k-mer and those of its revers complement, are almost equal. GCT, can be made more formal, as follows: the expected number of a DNA segment (k-mer) of length k, $w = (a_1 \dots a_k)$, indicated as $\langle n(w) \rangle$, and the expected number of its reverse complement $\langle n(\overline{w}) \rangle$, $\overline{w} = (\overline{a}_k, \overline{a}_{k-1}, \dots, \overline{a}_1)$ are equal. Let *N* be the total length of the duplex DNA. By definition of the expectation value, we have that:

$$\langle n(w) \rangle = \langle n(a_1, \dots a_k) \rangle = \sum_{i=1}^{N-k+1} P(X_i = w) = \sum_{i=1}^{N-k+1} P(a_i \dots a_{i+k-1}) \tag{8}$$

where $X_i = (a_i \dots a_{i+k-1})$ indicates the segment of the DNA sequence $X$ from position $i$ to position $i + k - 1$. The equality of $\langle n(w) \rangle$, and $\langle n(\overline{w}) \rangle$, is a consequence of the equality of the probabilities $P(X_i = w) = P(X_{N-i-k+2} = \overline{w})$. This follows by computing the average of the occurrence of the event $X_i = w$ with the probability distribution (7) for a generic position $i$ in the DNA sequence as:

$$P(X_i = w) = \frac{1}{Z} \sum_{X'} \delta(X_i', w) e^{-\beta \hat{H}(X')} \tag{9}$$

where $\delta(X_i', w) = 1(0)$ if $X_i' = w$ ($X_i' \neq w$). Since we sum over all possible sequences $X'$, for a generic function $f(X_i)$, we have that the identity $\sum_{X'} f(X') = \sum_{X'} f(\overline{X}')$ holds. Thus, we have:

$$P(X_i = w) = \frac{1}{Z} \sum_{X'} \delta(X_i', w) e^{-\beta \hat{H}(\overline{X}')} = \frac{1}{Z} \sum_{X'} \delta(X_{N-1-k+2}', \overline{w}) e^{-\beta \hat{H}(X')} = P(X_{N-i-k+2} = \overline{w}) \tag{10}$$

Using equation (8) we finally obtain the desired result:

$$\langle n(w) \rangle = \langle n(\overline{w}) \rangle \tag{11}$$

which is exact if the average is done over all the possible duplex DNA sequences. However, for a single duplex DNA of finite length *N*, corresponding to the analysed cases, equation (11) is only approximate and becomes exact in the $N \rightarrow \infty$ limit. This result proves the GCT. Furthermore, this also furnishes the explanation of the original Chargaff's second parity rule such as:

$$\langle n(a) \rangle = \langle n(\overline{a}) \rangle \quad \forall\, a \in \{A, C, G, T\} \tag{12}$$

According to our derivation, the "exceptional symmetry" found in the same strand of natural duplex DNA[2], is the most probable outcome that we can predict by chance. Of course, this does not mean that it must be always the case, but only that the Chargaff's rule is the most probable solution that we should expect when no other relevant things happen such as new constraints, besides to eq. (1).

## 1.3 Symmetry Predictions for k-mers of Arbitrary length Using the Maximum Entropy Principle

From the application of the maximum entropy principle we predict that, at equilibrium in long dsDNA, the numbers of k-mers and their reverse complements tend to be the same. However, other kinds of sequence symmetries, such as k-mers and their simple complement k-mers, are predicted not to hold. Here we study two completely different cases: $i$) the C-pairs of k-mers, $w = (a_1, a_{2,} \dots a_k)$ and their complements $w^c = (\bar{a}_1, \bar{a}_2, \dots \bar{a}_k)$, $ii$) RC-pairs of k-mers, $w$ and their reverse complements, $\overline{w} = (\bar{a}_k, \bar{a}_{k-1}, \dots \bar{a}_1)$. We can compute the conditional probability to find the k-mer $\overline{w}$ at position $j$ given that there is a k-mer $w$ at position $i$ and the conditional probability to find the k-mer $w^c$ at position $j$ given that there is a k-mer $w$ at position $i$:

$$P(X_j = \overline{w} | X_i = w) = \frac{\sum_X P(X)\delta(X_i, w)\delta(X_j, \overline{w})}{\sum_X P(\mathrm{X})\delta(\mathrm{X}_i, w)} \tag{13}$$

and

$$P(X_j = w^c | X_i = w) = \frac{\sum_X P(X)\delta(X_i, w)\delta(X_j, w^c)}{\sum_X P(\mathrm{X})\delta(\mathrm{X}_i, w)} \tag{14}$$

By summing over $i$ and $j$ in the equations (13) and (14) we obtain the expected number of C-pairs ($\langle n(w, w^c, k)\rangle$) and RC-pairs ($\langle n(w, \overline{w}, k)\rangle$):

$$\begin{aligned} \langle n(w, \overline{w}, k)\rangle &\equiv \frac{1}{2}\sum_{i,j=1}^{n(k)} \mathrm{P}(X_j = \overline{w}_i | X_i = w_i) \\ \langle n(w, w^c, k)\rangle &\equiv \frac{1}{2}\sum_{i,j=1}^{n(k)} \mathrm{P}(X_j = {w^C}_i | X_i = w_i) \end{aligned} \tag{15}$$

where $n(k)$ is the number of possible k-mers of size $k, n(k) = N - k + 1$. We can estimate the two sums introduced above by evaluating the expected value of the pairs of k-mers along a naturally duplex genome sequence of length $N$. Concerning the evaluation of $\langle n(w, w^c, k)\rangle$, the maximum entropy solution does not pose constraints to the C-pairs, thus assuming that the four bases occur in a completely uncorrelated way, the probability $\mathrm{P}(X_j = w^c | X_i = w)$ simplifies as $\mathrm{P}(X_j = w^c | X_i = w) = \mathrm{P}(X_j = w^c) \simeq \prod_a f_a^{n(a)}$, where $f_a$ is the frequency on the base $a$ in the DNA strand and $n(a)$ is the number of time the base $a$ appears in the k-mer $w^c$. With this assumption, the probability decreases exponentially with the length $k$. If we assume equal probability for all the bases, the probability of the k-mer becomes $P(w^c) \simeq 4^{-k}$, and the expectation can be estimated as:

$$\langle n(w, w^c, k)\rangle \simeq \frac{1}{2} n(k)^2 \cdot 4^{-k} \tag{16}$$

On the contrary, in naturally duplex DNA, the maximum entropy solution implies that the number RC-pairs has the same chance to appear, since the double helix constraint imposes that $\langle n(\overline{w})\rangle \simeq \langle n(w)\rangle$, form eq.(11). To find an analytical solution for $\langle n(w, \overline{w}, k)\rangle$, we observe that the frequencies of the k-mers and their reverse complments are predicted to be the same for long sequences ( $n(k) = N - k + 1 \gg 1$). Thus, we can assume that the square of the difference of the frequencies ($f(w) = n(w)/n(k)$) goes to zero for long DNA sequences (large $n(k)$), as:

$$\frac{1}{2}\sum_{w}\left(f(w)-f(\overline{w})\right)^2 \simeq \frac{1}{n(k)} \qquad (17)$$

From the equation (17) and the definition of the frequency, we obtain a simple formula for the number of RC-pairs given in equation (15), that is:

$$\langle n(w,\overline{w},k)\rangle \simeq \frac{1}{2}\left(\sum_{w} n(w)n(\overline{w})\right) \simeq \frac{1}{2}\left(\sum_{w} n(w)^2 - n(k)\right) = \frac{1}{2}\left(\sum_{i=1}^{n(k)} \langle n(w_i)\rangle - n(k)\right) \qquad (18)$$

## 1.4 Maximum Entropy Principle in single-stranded DNA (ssDNA)

In the case of ssDNA, the interaction energy depends only in the single chain $\widehat{H}(X)$, so that the average energy is:

$$\langle E\rangle = \sum_{X} H(X)P(X) \qquad (19)$$

Notice $H(X) \neq H(\overline{X})$. Similarly to what computed above we obtain:

$$P(X) = \frac{e^{-\beta H(X)}}{Z} \neq \frac{e^{-\beta H(\overline{X})}}{Z} = P(\overline{X}) \qquad (20)$$

Since in ssDNA $P(X) \neq P(\overline{X})$ there are no constraints applicable as in the dsDNA, so that the compositions and sequence regularities depend on the type of single-stranded energy $H(X)$ and the "temperature" factor $\beta$. However, we can predict that there is not a reason to expect that ssDNA genomes follow GCT.

## 2. TABLE OF NUCLEOTIDE RATIOS COMPUTED FOR DIFFERENT ORGANISMS

We extended and replicated previous observations[3,4] of Chargaff's second parity rule for 19566 dsDNA and 1397 ssDNA (single-stranded DNA) genomes including human chromosomes, bacteria, archaea, plastids, mitochondria, plasmids and viruses. In agreement with previous studies[3,4], we found strong Pearson correlations between the abundance of adenine versus thymine and cytosine versus guanine for human chromosomes, bacteria, archaea, plastids, plasmids and dsDNA viruses, and poor Pearson correlations for mitochondrial DNA (mtDNA) and ssDNA virus genomes (Table 1S).

**Table 1S – Frequency of nucleotides in different groups of organism genomes.**

| Organisms | Genome size average | Number of Organisms | $R^2$ (A vs T) | $R^2$ (C v G) |
|---|---|---|---|---|
| *Bacteria* | 3,584,631 | 1843 | 0.99 | 0.99 |
| *Archaea* | 2,010,000 | 170 | 0.99 | 0.99 |
| *Plastids* | 154,187 | 3004 | 0.82 | 0.88 |
| *Bacteria plasmids* | 55,960 | 2480 | 0.80 | 0.84 |
| *Archaea plasmids* | 122410 | 104 | 0.71 | 0.89 |
| *dsDNA viruses* | 44937 | 2911 | 0.63 | 0.73 |
| *Mitochondria* | 16566 | 9054 | 0.04 | 0.04 |
| *ssDNA viruses* | 2657 | 1397 | 0.02 | 0.10 |

*The $R^2$ is the Pearson correlation coefficient between the abundance of adenine versus thymine and cytosine versus guanine*

## 3. K-MERS VS REVERSE COMPLEMENT SEQUENCES ANALYSIS

GCT does not pose any spatial constraints to the relative positions of RC-pairs sequences. This is shown in Figure 1S and provides further support to our theoretical model. The genomic data were plotted using

LASTZ[5], a program for aligning entire genomes. Thus, we aligned k-mers versus their reverse complement sequences in different bacteria genomes and we found that the RC-pairs are equally distributed along the bacteria circular dsDNA. When eukaryote chromosomes (which are more competent for events involving recombination) were studied using LASTZ, we found interesting deviations from the expectations (left side of Figure 2S). However, when repeated regions (transposons, tandem repeats, low complexity regions, etc.) were removed (right side of Figure 2S), these deviations disappeared and eukaryotic regions rensembled prokaryotic genomes. This may seem to point, in eukaryotes, to a hybrid biological/physical model, indicating that it is possible to define two portions in these genomes: one that similarly to the prokaryote at the Chargaff's equilibrium, whereas a second portion consists of repeated regions that originated from recent duplication events and that are either still evolving towards the equilibrium or are kept as such by biological selective pressures.

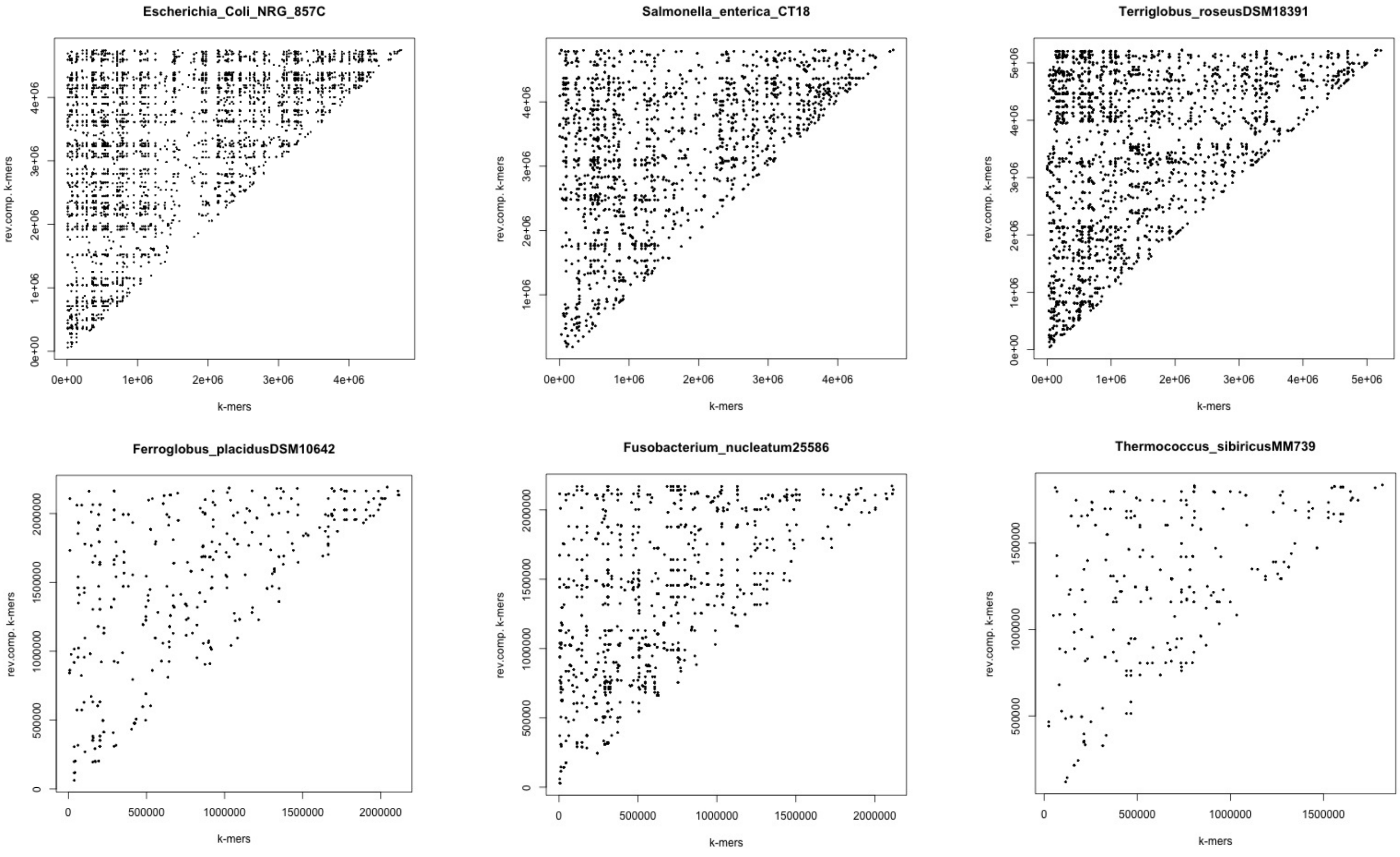

**Figure 1S | Scatterplot of bacteria k-mers vs their reverse complement sequences**. These plots show six bacteria k-mers against their reverse complement sequences. Each point in the x-axis represents the position of the first base of a specific k-mer, whereas each value within the y-axis is the first base of the reversed complement sequence. The six plots show that the RC-pairs are uniformly distributed all along different bacteria genomes, which is in agreement with our maximum entropy based prediction.

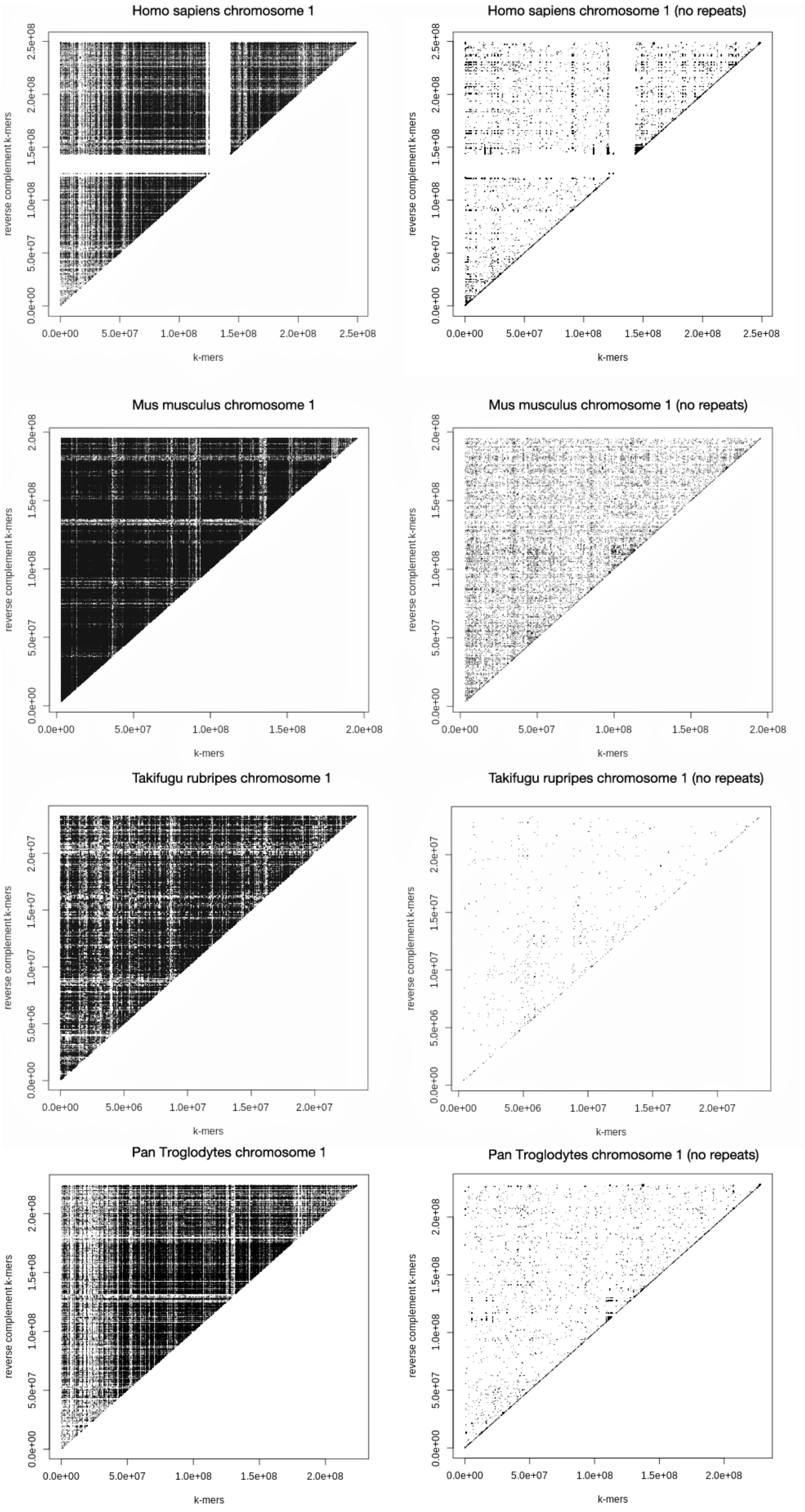

**Figure 2S | Scatterplot of k-mers against their reverse complement in six eukaryotic chromosomes with and without repeated regions.** The presence of a point of coordinates *(x, y)* shows the occurrence of a RC-pair where x represents the position of the first base of a specific k-mer, whereas the y is the first base of its reversed complement sequence along a single strand of eukaryotic genomes. RC-pairs tend to be enriched in specific regions (genomes on the left side of the figure) of the eukaryotic genomes, consistent with being the result of a sequence rearrangement or recombination (region not at equilibrium). However, when the repeated regions were removed from eukaryotic genomes (right side of the figure) the points are uniformly distributed (region at equilibrium) like in prokaryotic dsDNA.

## 4. CODON TRIMERS FREQUENCY BARPLOT

The human coding regions are constrained by the fact that trimer codons are translated into amino acids following a species-specific code. Thus, we investigated whether, even in a given codon bias system, the GCT is still detectable in human codon usage. A positive correlation was found when analysing the frequency of the codons and the frequency of the corresponding reverse-complement codons (Pearson correlation=0.5 with a p-value < $10^{-4}$) indicating that, still, there is a significant CGT trace in the human codon usage. The trimer frequency at whole genome level perfectly follows the GCT as shown in Figure 3S.

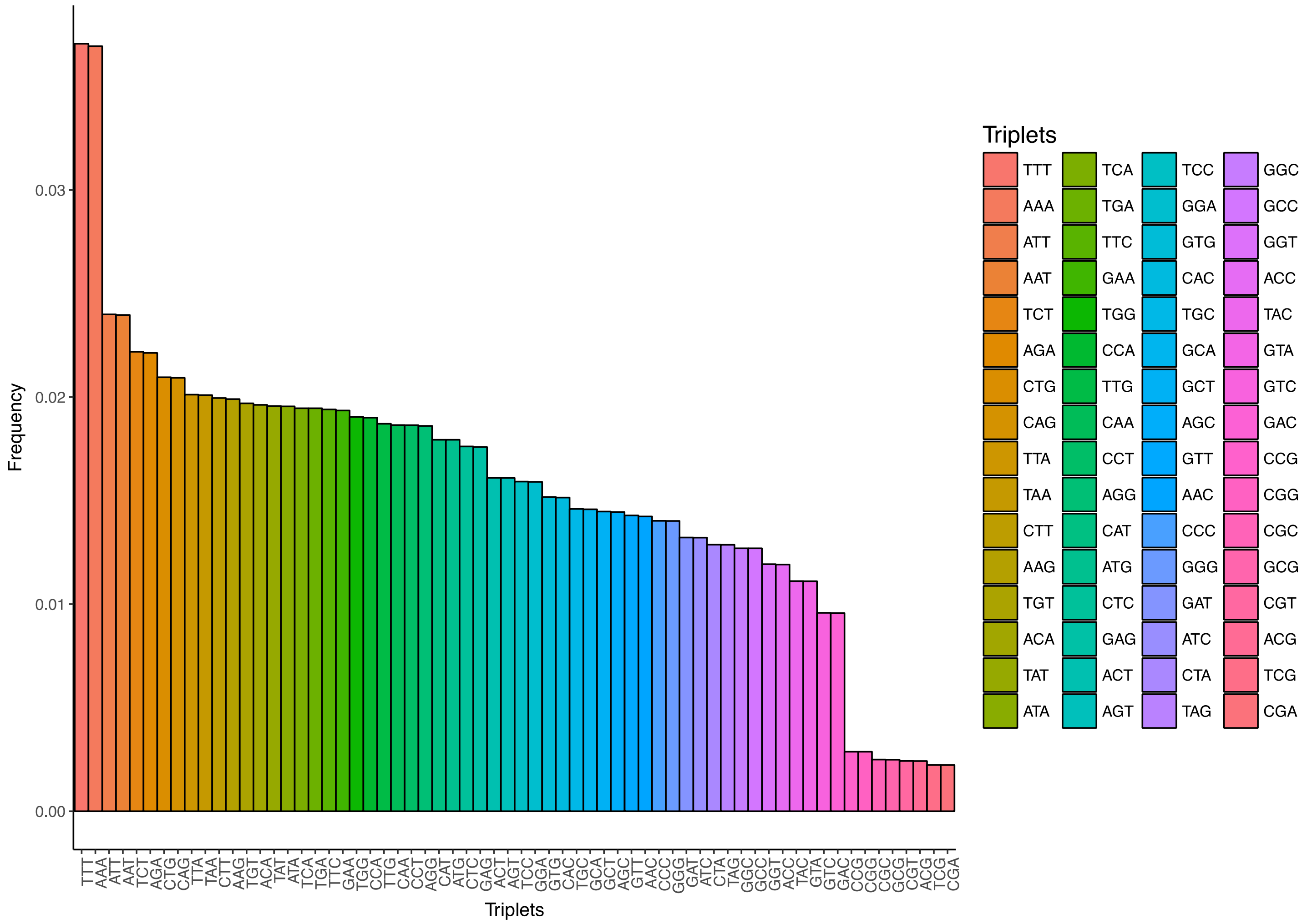

**Figure 3S |** Trimer as a function of their frequency (y-axis) in the human genome. The RC-pairs sit nearby as predicted by the Generalized Chargaff Theory. For the sake of clarity the trimers are highlighted in coluour on the right side

## 5. ANALYSIS OF CONTEXT-DEPENDENT MUTATIONS PATTERNS IN HUMAN POPULATION

Unbalanced distribution of context dependent mutations has been reported for the human genome when looking at trimers (considering the mutating nucleotide, the one immediately upstream and the one immediately downstream)[6] and at longer k-mers[7]. Here we speculate that such unbalances may be partly explained by GCT and maximum entropy, given that each new mutation, regardless of its biological impact, may be under selective constraints for its contribution towards the global energy equilibrium. To investigate this possibility, we studied 100 European (CEU) whole genome sequences from the 1000 Genome Project[8] and focussed on N**A**N -> N**B**N mutations, where A is always the ancestral (Consensus ancestral sequence from the 1000 Genomes Project, B is the derived allele and N are any two nucleotides defining the specific sequence context of a given A/B mutation. We described each of the 192 possible N**A**N -> N**B**N mutation types by using the following measures: i) abundance along the genome of the mutation event (the number of times a given mutation is observed to have happened, along the genome, even just in a single individual) and ii) average population frequency of the mutations (the average fraction of individuals carrying a given mutation, calculated along the whole genome). The first measure recapitulates the properties of the human genome in terms of sequence content and, hence, the opportunity for a given N**A**N trimer to mutate to N**B**N (the more the occurrences of N**A**N along the ancestral genome sequence, the higher the chances of observing a N**A**N -> N**B**N mutation event). The second measure is independent from the properties of the ancestral genome sequence and is instead a measure of the evolutionary success (or spread across the population) of a given mutation event (under the simplifying assumption of no recurrent mutations, and identity by descent of all carriers of a given mutation at a given genomic site). By scattering these two mesures (**Figure 4Sa**) we noticed that all 192 N**A**N -> N**B**N mutation types fell into three distinct categories: i) Low overall occurrencies, intermediate average population frequencies (Class 1); ii) High overall occurrencies, high average population frequencies (Class 2); iii) High overall occurrencies, low average population frequencies (Class 3). Somewhat expectedly from Figure 4Sa, the number of occurrencies of a given N**A**N -> N**B**N substitution is directly proportional to the one of its N**B**N -> N**A**N reverse ($R^2$=0.89, p=$10^{-16}$), while its average population frequency is inversely proportional to the average frequency of the reverse substitution ($R^2$=-0.75, p=$10^{-16}$). We also noticed that Class 1 mutations include only transversions (purine <-> pyrimidine substitutions), while Class 2 and Class 3 mutations include only transitions (purine-purine or pyrimidine-pyrimidine substitutions). In addition, Class 2 only include G -> A and C -> T, while Class 3 contains A -> G and T -> C mutations, suggesting a role of the directionality of these mutation types within the broader genomic "ecosystem". We then inspected the realtionship between the average population frequencies of a mutation and the one of its RC-pair (pairs involved in the GCT, here defined, for a given N**A**N -> N**B**N, as mutations substituting the reverse complement of N**A**N into the reverse complement of N**B**N) and discovered a remarkably good correlation ($R^2$=0.91, p=$10^{-16}$, **Figure 4Sb**). Considering the average frequency of a given mutation type within the European population as a measure of its evolutionary success (or of its age), we then concluded that the evolutionary trajectory of any N**A**N -> N**B**N mutation is paired with the trajectory of mutations substituting the reverse complement of N**A**N into the reverse complement of N**B**N, once more confirming predictions based on GCT and maximum entropy equilibrium.

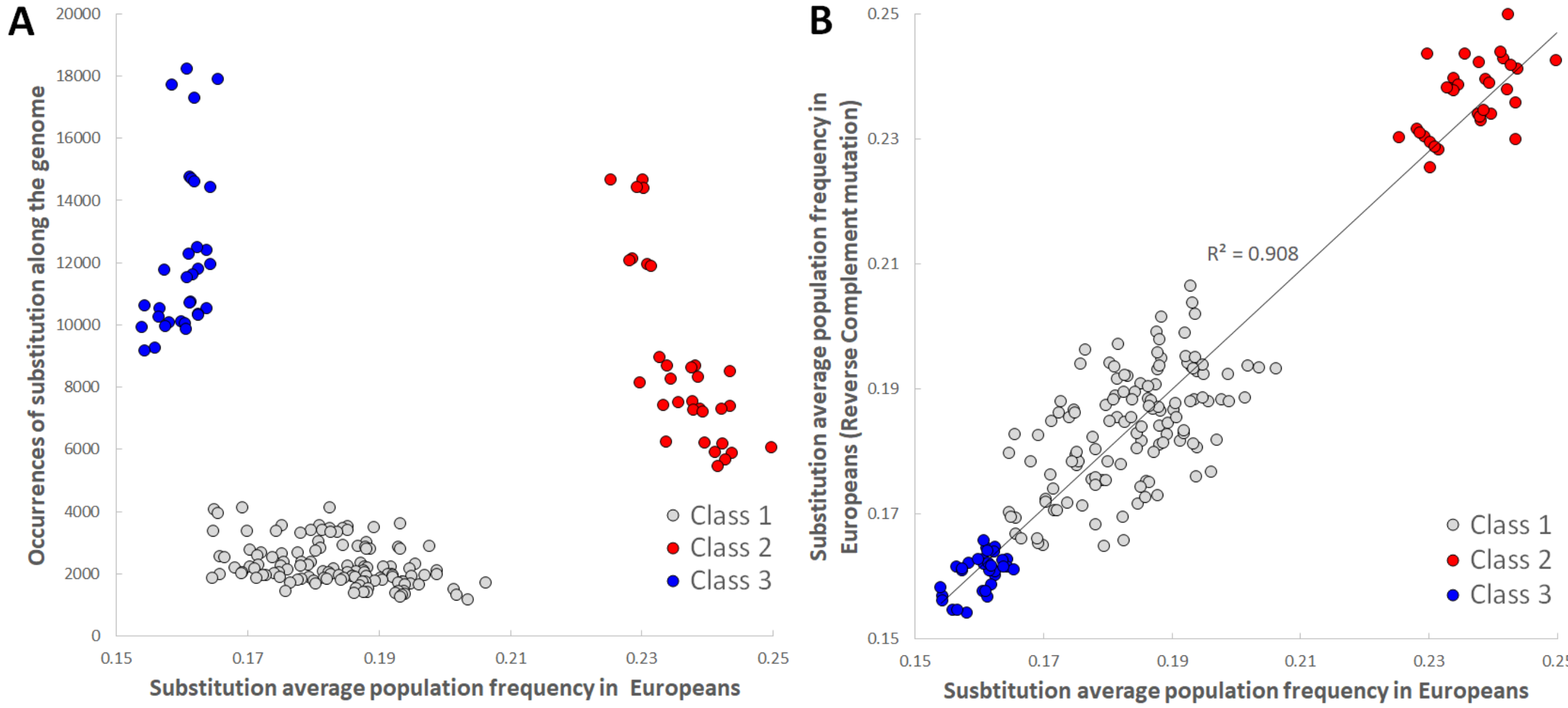

**Figure 4S. Context dependent substitutions within the human genome (1000 Genomes Project European samples). a,** The average population frequency of a given N**A**N->N**B**N substitution (x-axis) and the total occurences of that substitution along the genome (y-axis) separate all the possible trimers into three distinct categories (Class 1, Class 2, and Class 3). **b,** The average population frequency of all the events of a given N**A**N->N**B**N type (x-axis), calculated on 1000 Genomes Europeans (CEU) correlates ($R^2$=0.91, p=$10^{-16}$) with the average population frequency of the reverse complement mutation type (y-axis: the mutation substituting the RC of N**A**N into the RC of N**B**N).